\documentclass{JINST}
\usepackage[dvips]{epsfig,graphicx}

\title{Charge collection in the Silicon Drift Detectors of the ALICE
  experiment}
\author{
B.~Alessandro$^a$, 
R.~Bala$^b$,
G.~Batigne$^{a,\circ}$, 
S.~Beol\'e$^b$,
E.~Biolcati$^b$\thanks{Corresponding author: biolcati@to.infn.it}, 
P.~Cerello$^a$, 
S.~Coli$^a$, 
Y.~Corrales~Morales$^{c,a}$, 
E.~Crescio$^a$,
P.~De~Remigis$^a$, 
D.~Falchieri$^d$, 
G.~Giraudo$^a$, 
P.~Giubellino$^a$, 
R.~Lea$^b$, 
A.~Marzari Chiesa$^b$, 
M.~Masera$^b$, 
G.~Mazza$^a$,
G.~Ortona$^b$,
F.~Prino$^a$, 
L.~Ramello$^e$, 
A.~Rashevsky$^f$, 
L.~Riccati$^a$, 
A.~Rivetti$^a$,
S.~Senyukov$^e$, 
M.~Siciliano$^b$, 
M.~Sitta$^e$,
M.~Subieta$^b$, 
L.~Toscano$^a$,
F.~Tosello$^a$\\
\llap{$^a$}Istituto Nazionale di Fisica Nucleare - Sezione di Torino, Torino, Italy\\
\llap{$^b$}Dipartimento di Fisica Sperimentale dell'Universit\`a di Torino 
and INFN, Torino, Italy\\
\llap{$^c$}Instituto Superior de Tecnolog\`ias y Ciencias Aplicadas (InSTEC), Havana, Cuba\\
\llap{$^d$}Dipartimento di Fisica dell'Universit\`a di Bologna 
\& INFN Bologna, Italy\\
\llap{$^e$}Dipartimento di Scienze e Tecnologie Avanzate
dell'Universit\`a del Piemonte Orientale \& Gruppo Collegato INFN, Alessandria, Italy\\
\llap{$^f$}Istituto Nazionale di Fisica Nucleare - Sezione di Trieste, Trieste, Italy\\
\llap{$^\circ$} now at SUBATECH, Ecole des Mines de Nantes, 
Universit\'e de Nantes, CNRS/IN2P3, Nantes, France\\
}
\abstract{A detailed study of charge collection efficiency has been performed on
the Silicon Drift Detectors (SDD) of the ALICE experiment. Three different methods to
study the collected charge as a function of the drift time have been implemented.
The first approach consists in measuring the charge at different injection
distances moving an infrared laser by means of micrometric step motors. The second method 
is based on the measurement of the charge injected by the laser at
fixed drift distance and varying the drift field, thus changing the
drift time. In the last method, the measurement of the charge deposited by
atmospheric muons is used to study the charge collection efficiency
as a function of the drift time.
The three methods gave consistent results and indicated that
no charge loss during the drift is observed for the sensor types used
in 99\% of the SDD modules mounted on the ALICE Inner Tracking System.
The atmospheric muons have also been used to
test the effect of the zero-suppression applied to reduce the data
size by erasing the counts in cells not passing the thresholds for noise removal.
As expected, the zero suppression introduces a dependence of the
reconstructed charge as a function of drift time because it cuts the
signal in the tails of the electron clouds enlarged by diffusion
effects. These measurements allowed also to validate the correction
for this effect extracted from detailed Monte Carlo simulations of the
detector response and applied in the offline data reconstruction.}

\keywords{dE/dx detectors; Particle tracking detectors}

\begin{document}
  
\section{Introduction}

Large area Silicon Drift Detectors (SDDs)~\cite{Gatti,dsi1,dsi2} equip the two
intermediate layers of the Inner Tracking System (ITS) of the ALICE experiment
at the LHC~\cite{PPR2,JINSTalice}. They have been selected due to their good
spatial resolution, capability of unambiguous two-dimensional position
determination and possibility to provide the energy-loss measurement needed for
particle identification.
The operating principle of SDDs is based on the drift towards collecting
anodes of the electrons produced in the
sensitive volume by an ionizing particle. The transport of electrons in a
direction parallel to the surface of the detector and along distances
of several centimeters is achieved by creating
a drift channel in the middle of the depleted bulk of a Silicon wafer,
as shown in figure~\ref{fig:sdd}.a. 
Thus, the distance of the crossing point from the anodes is determined by the
measurement of the drift time, as long as the drift velocity is know, which is
proportional to the applied electric field $E$ and to the electron mobility
$\mu_{\rm e}$ ($v=\mu_{\rm e} E$). The second coordinate is obtained from the
centroid of the charge distribution along the anodes.

To reach the required spatial resolution of $\approx$30~$\mu$m,
either an excellent uniformity of the drift field over all the sensitive
region of the detector, or to correct for the systematic errors caused by its
non-uniformity is necessary, as discussed in \cite{articolomappature}. 
Furthermore, the drift
velocity must be known with a precision better than 0.1\% in every point of
the SDD sensors. This is a challenging requirement because the mobility depends
on the temperature as $\mu_{\rm e}\propto T(K)^{-2.4}$, so the drift speed, which is
about 6.5~$\mu$m/ns at the bias voltage of 1.8~kV, varies 
by about 0.8\%/K at room temperature. 

In order to perform the study of the collection charge efficiency in the SDD,
a test station has been set-up at the INFN Technological Laboratory in Turin. In
the following sub-sections an overview of the ALICE SDDs and the explanation of 
the methods used for the data analysis will be presented. In section~\ref{sec:laser}
the results obtained by using the infrared laser with two different data taking
procedures will be reported. In section~\ref{sec:cosmic} the study performed with
cosmic data will be presented and the results will be discussed. The conclusions of this 
work are addressed in the last section (\S~\ref{sec:conclusion}). 

\subsection{The ALICE Silicon Drift Detector}

\begin{figure}[!ht]
  \centering
  \includegraphics[width=0.49\textwidth]{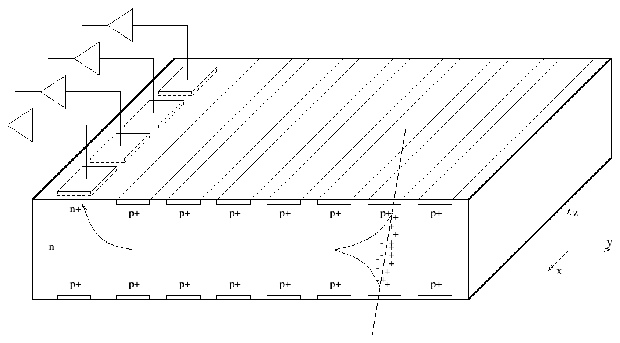}\ \ 
  \includegraphics[width=0.49\textwidth,height=5cm]{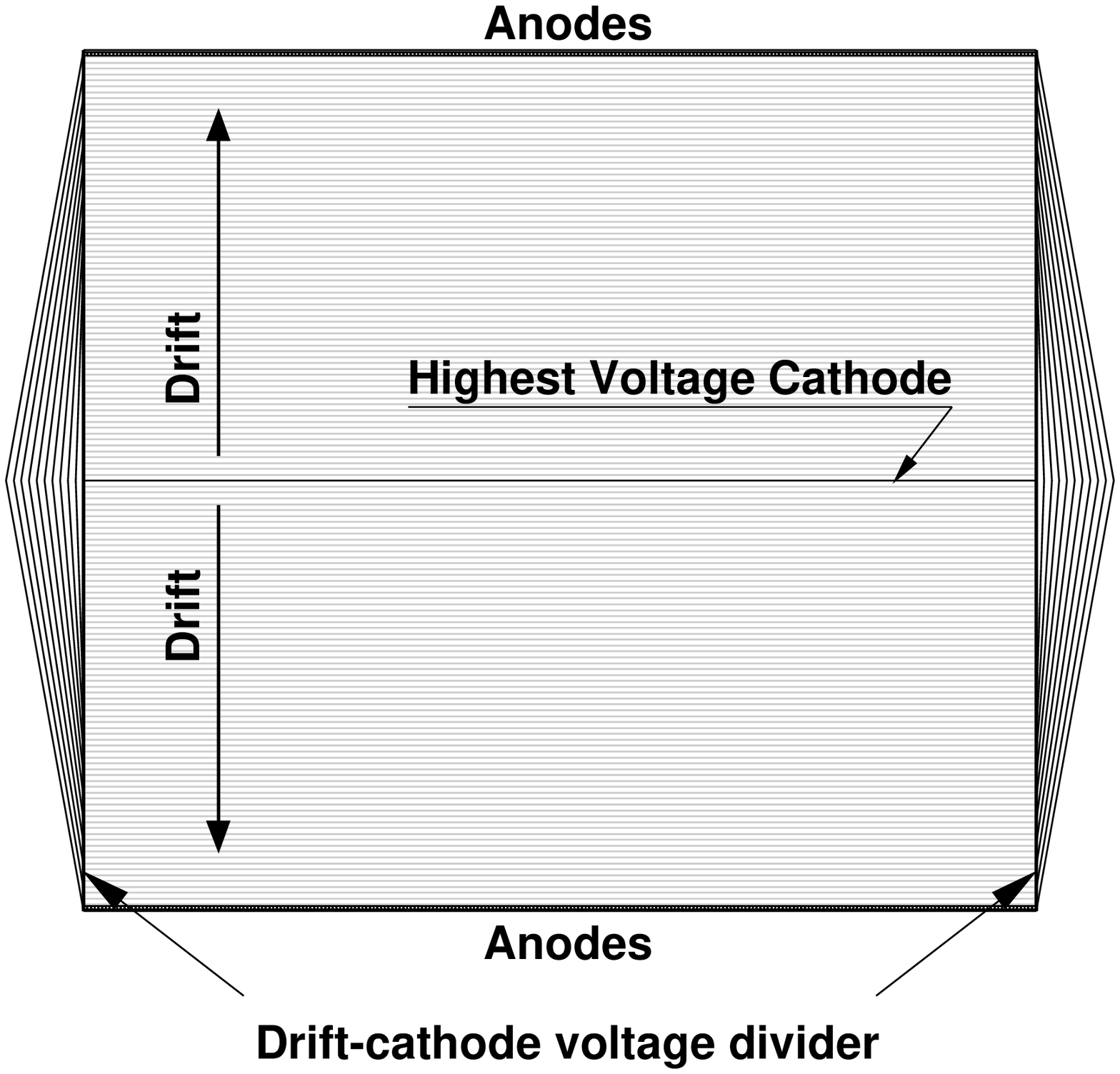}\\
  (a) \hspace{7cm} (b)
  \caption{(a) SDD operation scheme. (b) SDD module.}
  \label{fig:sdd}
\end{figure}

\paragraph{The sensor}
The SDD sensors~\cite{sasha} of the ALICE experiment are built on n-type high-resistivity
300~$\mu$m-thick Neutron-Transmutation-Doped silicon.
The active area of 7.02~$\times$~7.53~cm$^{2}$ is split into two
drift regions ($\approx$~35~mm long) by a central cathode strip which is biased
at a maximum voltage (HV) of -1700 $\div$ -2400 V. In each drift region, on both the
detector surfaces, 291 p$^+$ cathode strips with 120 $\mu$m pitch are implanted
as sketched in figure~\ref{fig:sdd}.b. A built-in voltage divider, made of
Polysilicon implants, biases these cathodes at a gradually decreasing voltage
from the HV applied to the central cathode down to a medium voltage
(MV$\approx$~-40 V) applied to the last cathodes before the anodes. This MV is
used to polarize the so-called collection region, as described in~\cite{articolomappature}.
In this way, a drift field parallel to the wafer surface is generated,
giving rise to a bi-directional structure: the electrons drift from the central
cathode towards the anodes. The drift field $E_{\rm drift}$ is given by the
ratio between the inter-cathode voltage drop $V_{\rm gap}=(HV-MV)/291$ and the
cathode pitch (=~120~$\mu$m); typical operation values for $V_{\rm gap}$ are
between 5.5 and 8 V, corresponding to $E_{\rm drift}$ in the 458-667~V/cm range.
At the end of each of the two drift regions, the electrons produced by the
crossing particle are conveyed by means of pull-up cathodes, placed
below the anodes, towards an array of 256 collection anodes (294~$\mu$m pitch)
connected via micro-cables to the front-end electronics. For a detailed
description of the SDD sensor, see ~\cite{JINSTalice,articolomappature,sasha}.

\paragraph{Front-end electronics and zero-suppression}\label{par:frontend}
The front-end electronics of the SDD is based on three application-specific
integrated circuits (ASICs). The first one, called PASCAL, is a mixed-mode chip
with 64 channels~\cite{Rivetti1}. In each channel the signal coming from one
anode is amplified by a charge sensitive amplifier and sampled at
40~MHz\footnote{A sampling frequency of 20~MHz can also be used in order
to reduce the dead time.} by an analogue memory with 256
cells. When a trigger signal is issued, the content of the memory is frozen and
the samples are digitized by a 10-bit successive-approximation ADC. One
converter serves two adjacent channels, so 32 ADCs are embedded on the chip.
After the digitization, the data are transferred to the second stage, handled by a
 AMBRA chip~\cite{Mazza1}. This ASIC performs the pedestal equalization on a
channel by channel basis  and applies a 10 to 8 bits compression algorithm
before  storing the data in one of its four event buffers. Four PASCAL-AMBRA pairs are
mounted on the front-end hybrid~\cite{Rivetti1}, which is a flex circuit made of
Aluminum-Kapton cables laid-out on a carbon fiber support. Two front-end hybrids
are hence necessary to read-out a full sensor. A short cable ($\approx$2~cm)
connects the SDD anodes to the PASCAL inputs, while a longer one (up to 40
cm) allows the communication between the hybrids and the rest of the system and
distributes the supply voltages. All the interconnections exploit the Aluminum on
Kapton technology.

Two front-end hybrids are connected to the same data compression board which hosts
one CARLOS chip~\cite{Bolognesi1}. The use of four event buffers on AMBRA allows the
derandomization of the triggers, so the data transmission speed from the
front-end hybrid can be tuned to the average event rate. CARLOS performs the
zero suppression before sending the data via optical fiber to the so
called CARLOSrx board~\cite{samuele}. Due to the
diffusion occurring in the sensor the signals present significant tails, so the
use of a simple zero suppression is problematic. A bi-dimensional compression
algorithm based on a dual threshold has therefore been preferred~\cite{AlbertW}.
To be accepted as a valid signal, a sample must exceed the higher threshold and
have at least one neighbor above the lower one or viceversa. This allows to
suppress noise spikes (isolated samples above threshold) and to preserve as much
as possible the samples in the tail of the signals. Despite their amplitude
these can in fact contribute significantly to the final spatial resolution
because of their bigger lever arm in the centroid calculation. Moreover, a cut
of these tails also affects the measurement of the deposited energy.

\paragraph{Charge collection}
The SDDs inside the ALICE ITS have two main tasks: the first is to ensure
an adequate space resolution on the particle crossing point together with good
multi-track capability, while the second is to measure the specific ionization
energy loss (dE/dx). Hence, the Charge Collection Efficiency (CCE) is an
important characteristic of the detector quality in order to achieve the
required precision in dE/dx measurements.
The CCE must also be known as a function of the drift time (i.e. the time necessary 
for charge carriers electrons or holes, created by ionizing particles in active volume,
to reach a signal readout electrode). In the SDD, the drift time can be as long as
6~microseconds. The possible reasons of a decrease in CCE are described in the
following points.
\begin{enumerate}
\item The drifting charge carriers (electrons in the SDD case) undergo
diffusion, giving rise to an electron cloud with Gaussian-like profile, both along
the drift and anodes axes, with a sigma given by 
  \begin{equation}
    \sigma^2 = 2\,D\,t_{\rm{drift}} + \sigma_{\rm{time0}}^2
  \end{equation}
  
  where $D$ is the diffusion coefficient: $D=K_{\rm B}T\mu_{\rm e}/q$,
with $K_{\rm B}$ the Boltzmann constant, $T$ the absolute temperature,
$\mu_{\rm e}$ the electron mobility and $q$ the electron charge. For the ALICE
Silicon Drift Detectors, $D\approx 3\div5\,\mu\rm{m}^2$/ns.

  The electron cloud generated far from the anodes can extend up to 4 anodes in
the anode direction and can last up to 200~ns along the drift direction. It may
well happen that a fraction of the charge in the tails of the electron cloud
does not contribute to the total collected charge, because it gets suppressed by
the zero-suppression algorithm, described in the previous
section. This fraction increases with the Gaussian width $\sigma$ and
consequently with increasing drift time. 
The zero-suppression may also affect the fraction of collected charge in case of
inclined tracks which give rise to elongated clusters with a larger fraction
of anode/time bin cells with signal below the thresholds. This effect is however not
present in these studies because particles orthogonal to the detector surface have
been used in both the laser and cosmic studies.
\item A localized defect in one of the voltage dividers could result in the voltage unbalance 
between corresponding drift cathodes placed on opposite SDD surfaces (see
figure~\ref{fig:sdd}.b). This effect leads to a shift of the bottom of the
potential gutter, along which the electrons drift towards the surface where
the charge can be trapped. As a consequence, it is expected to have a \emph{step
like} CCE fall-down above a given drift distance.
\item Impurities present in the depleted silicon bulk of the SDD could trap
drifting charge carriers which causes a dependence of the collected charge 
on the carriers drift time. It is worthwhile to note that the
effect of the zero-suppression algorithm mentioned in comma 1) easily fakes
charge trapping.  
\end{enumerate}

\subsection{Experimental setup and analysis methods} \label{par:method}
The test setup exploits an infrared laser and micro-metric step motors to
provide the capability of generating signals in known positions in the detector.
For a detailed description, see~\cite{articolomappature}. The aim of this work
is to study the dependence of the collected charge on the drift time, so as to test
the possible presence of systematic effects on the CCE.
The studies have been performed on three Silicon Drift Detectors that were not
mounted on the ALICE ITS, because they present a large number of bad
channels (noisy or non-functional), but they can be used for this analysis that
is performed on few selected anodes. Two modules (called A and B) have been obtained from the final
production of sensor using Silicon wafers with a uniform dopant
concentration. One of these (B) has a localized defect in the internal voltage
divider. The third module (called C) was a prototype built from a different Silicon wafer and
presents significant doping inhomogeneities~\cite{articolomappature}.

Three different analysis methods have been implemented: two of them make use of
the 980~nm infrared laser to generate the signal, while the third method is
based on the ionization produced in the SDD sensor by atmospheric muons. 
\begin{enumerate}
  \item \emph{Fine position scanning} consists in measuring the collected charge as a
function of the drift distance (i.e. the drift time), moving the laser on a
linear trajectory along the drift coordinate at fixed anode coordinate by means of
micro-metric step motors.
  \item \emph{Fixed position} is based on the measurement of the collected
charge when firing the laser in a fixed point (i.e. at fixed drift distance) and
varying the drift field, thus changing the drift time only. 
  \item \emph{Cosmic rays} is based on the measurement of the charge deposited by
atmospheric muons as a function of drift time.
\end{enumerate}

\begin{figure}[!ht]
  \centering
  \includegraphics[width=0.49\textwidth]{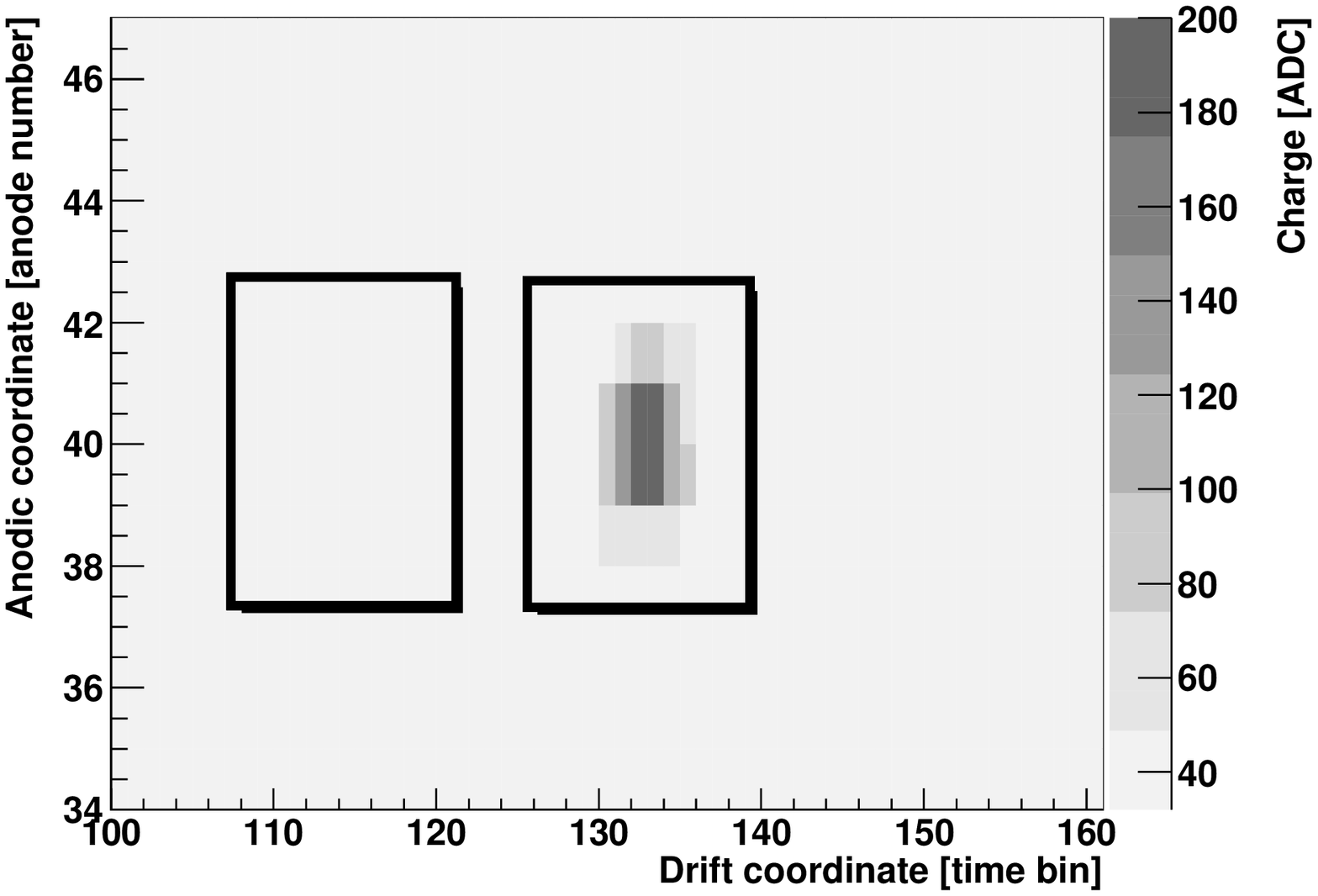}\ \ 
  \includegraphics[width=0.49\textwidth]{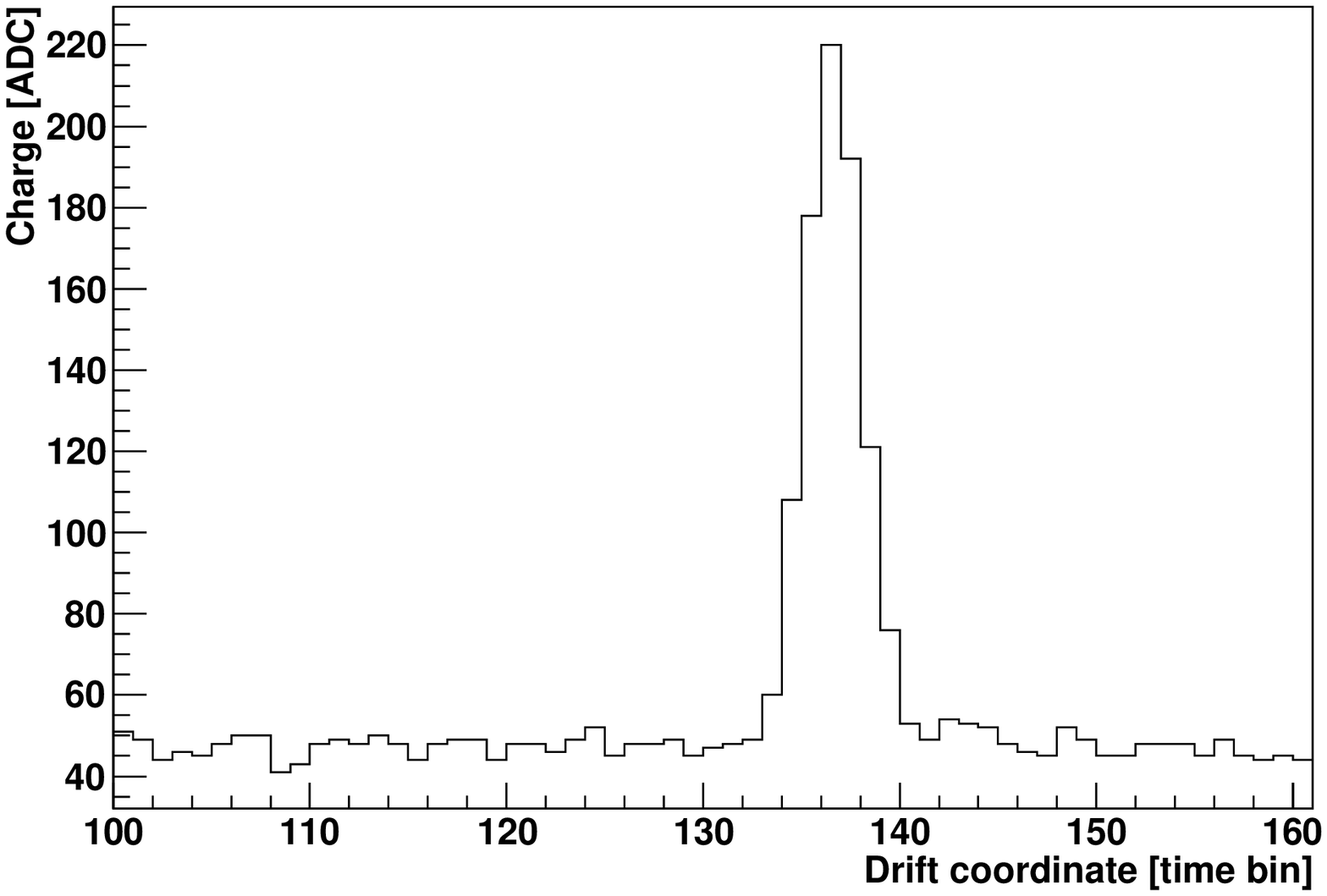} \ \ 
  (a) \hspace{7cm} (b)
  \caption{(a) Cluster generated by a laser shot. The total charge value is
obtained by subtraction of charge contained in the two rectangles. (b) Slice
plot along $x-$axis: charge collected by the anode corresponding to the signal
peak as a function of the drift time.}
  \label{fig:peak1}
\end{figure}

Data have been collected without enabling the zero-suppression algorithm to
isolate possible effects of charge trapping and voltage divider defects.
In figure~\ref{fig:peak1}.a an example of a cluster produced by a laser shot in the middle between anodes 39 and 40 is shown.
On the $x-$axis the drift time measured in time bins (1~time bin~=~25~ns)
is reported, while on the $y-$axis the anodic coordinate, defined  by the anode
number, is shown. The collected charge, expressed in ADC counts,
is represented in gray scale. The two rectangles represent the \emph{signal} and the
\emph{baseline} regions which are used in this analysis to extract the collected
charge, as it will be explained later in this section. The plot shows 60~time
bins $\times$ 17~anodes (out of 256~$\times$~256 cells of one hybrid) where the
signal \emph{cluster} is visible. The cluster is 6~time bins $\times$ 4~anodes
size, with a peak value of $\approx$~200~ADC. 
The size of the cluster is not due to the size of the laser spot (which is $\approx$~5$\div$10~$\mu$m, but to charge diffusion effects. In the remaining part of the
sensor an average value of $\approx$~40~ADC is measured for the baseline.
In figure~\ref{fig:peak1}.b the charge collected by the anode corresponding to
the signal peak is shown as a function of the drift coordinate. For this module, the average
noise value (i.e. the fluctuation around the baseline) is
$\approx$2.3~ADC counts. The peak signal-to-\emph{noise} ratio, obtained after the
baseline subtraction, is about~95. 

 A possible method to measure the total cluster charge would be to set a
threshold equal to the baseline increased by few times the noise, and sum the
ADC counts of the cells passing this threshold (somewhat equivalent to a
zero-suppression algorithm). Anyway, with such on approach, the tails of
the cluster would be cut thus affecting the measurement of the charge
especially in the cases of large drift times. 

Therefore a different method has been developed. The total charge of the cluster
is obtained by summing the ADC counts of all the cells inside a rectangular
region centered on the signal peak and sized so as to contain the entire
cluster, also for the cases with maximal diffusion (i.e. largest drift time).
The sum of the ADC counts in a rectangle with the same area and shifted along
the drift direction is subtracted to remove the contribution of the baseline
under the peak. A sketch of the two regions, called \emph{signal rectangle} and
\emph{baseline rectangle}, is shown in figure~\ref{fig:peak1}.a.

\section{Laser measurements}\label{sec:laser}

\subsection{Fine position scanning} \label{sec:moduleab}
To check the dependence of the collected charge on the drift distance, a
specific trajectory has been implemented in the motor controller, which moves
the laser along the drift direction (i.e. perpendicular to the collection anode
row) at a fixed anode coordinate.

\begin{figure}[!ht]
  \centering
  \includegraphics[width=0.39\textwidth,height=4.6cm]{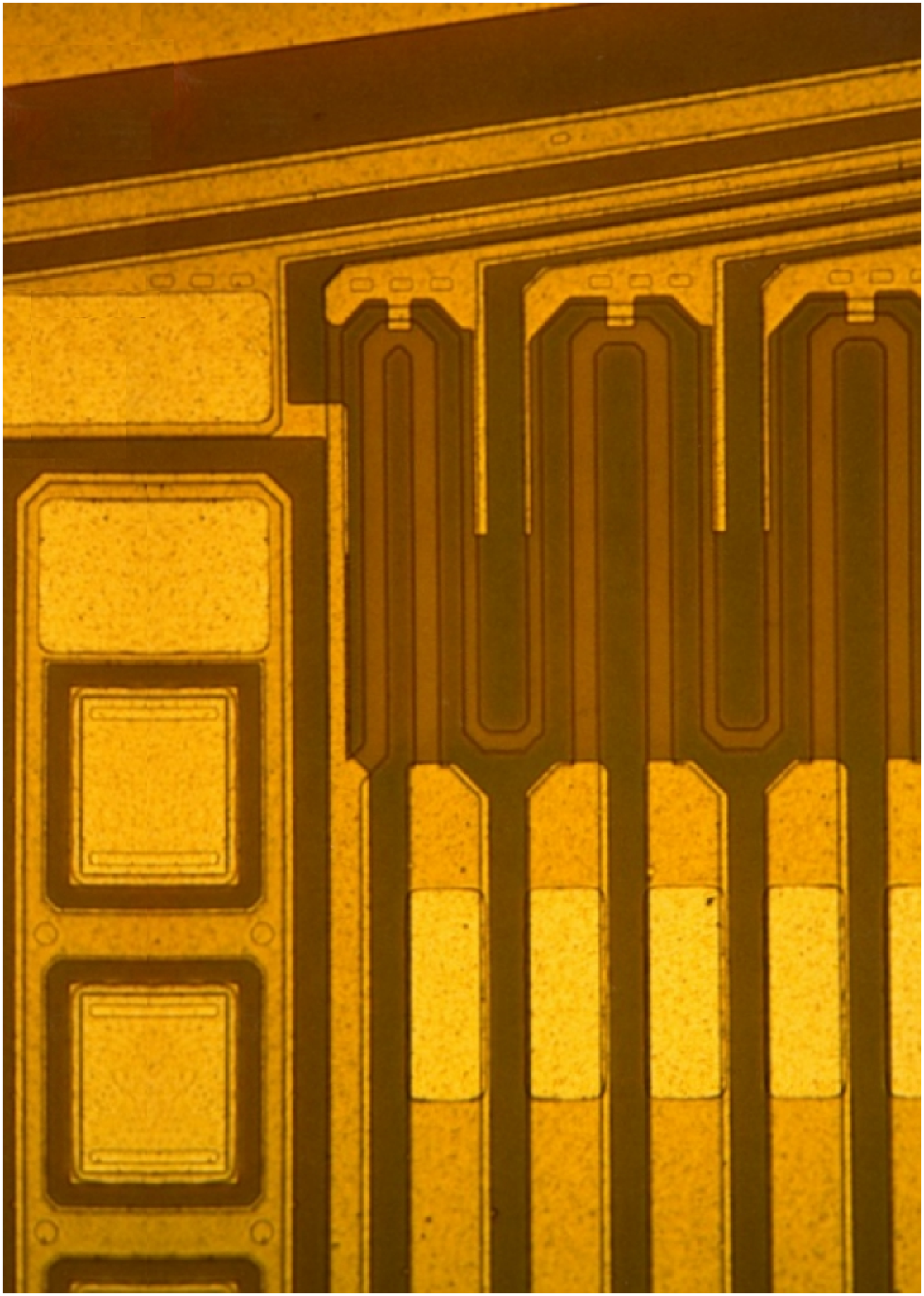} 
  \includegraphics[width=0.59\textwidth,height=5cm]{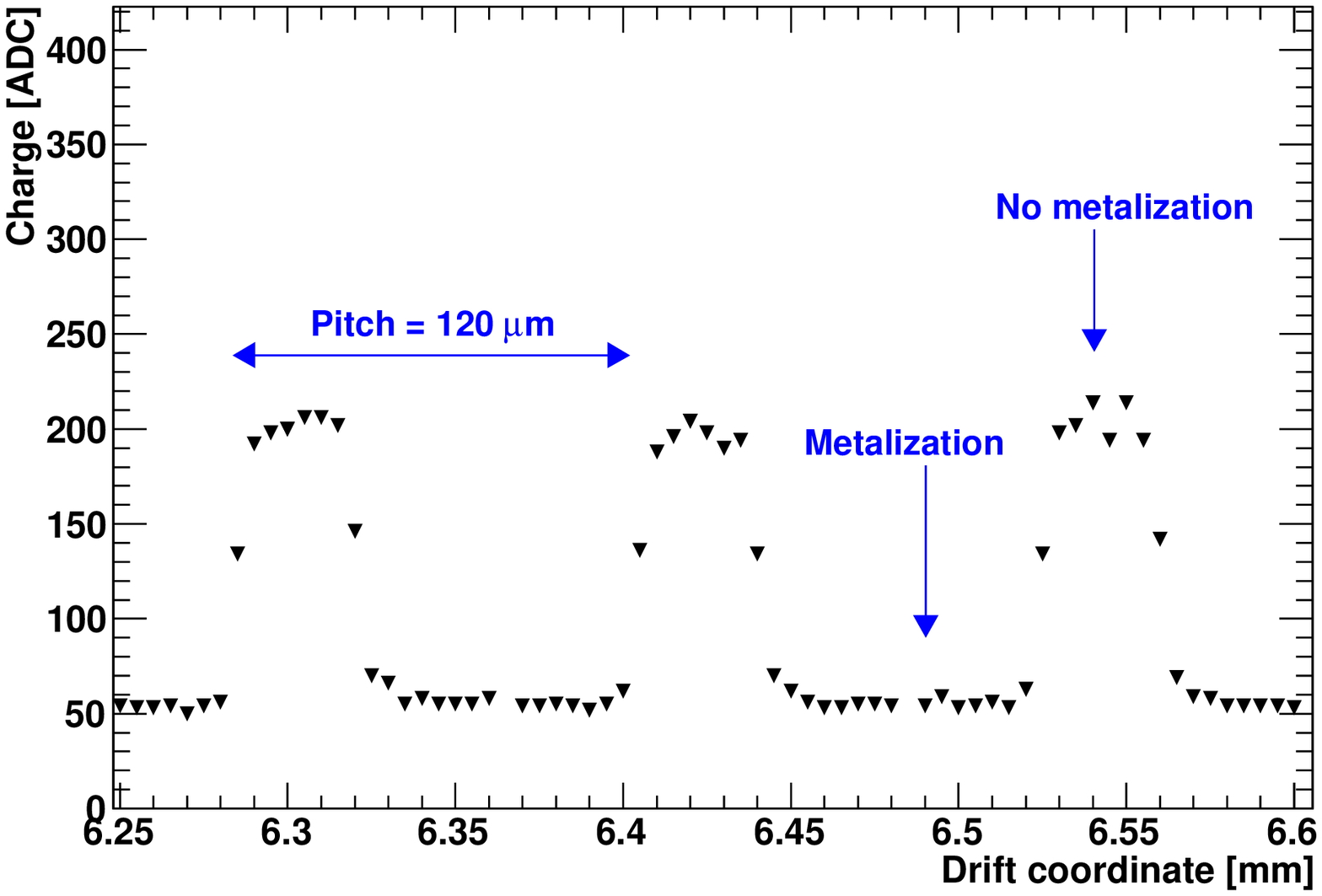}\\
  (a) \hspace{6cm} (b)
  \caption{(a) Picture of the SDD sensors (zoom): the cathode strips, the
voltage divider (at the top) and the collection anodes (on the left) are
visible. (b) The metal/oxide pattern of the SDD module visible from the laser
signal peak vs drift coordinate.}
  \label{fig:pattern}
\end{figure}

The SDD module~\cite{sasha} has a row of 291~cathode strips perpendicular to the
drift direction. A 85~$\mu$m wide metallization, which reflects the laser,
covers the central part of each cathode (figure~\ref{fig:pattern}.a). In order 
to properly generate a signal it is therefore necessary to center the laser spot on
the 35~$\mu$m wide space between two Aluminum strips where no metal is present. 
A trajectory with a spatial gap of 5~$\mu$m between two consecutive laser shots
has been implemented, to allow to select those in which the laser photons have not been reflected by the
metallization. In figure~\ref{fig:pattern}.b the value of charge peak (i.e. the
ADC counts in the anode/time~bin cell with highest signal) as a function of the
laser position along the drift direction is plotted. It is possible to
distinguish the metal/oxide pattern (pitch = 120~$\mu$m) of the SDD sensor: the
flat regions in which the signal peak is low (50~ADC counts, close to the
baseline value) correspond to the metallization while in the inter-cathode
regions a peak signal of $\approx$~200~ADC counts is observed.

\begin{figure}[!ht]
  \centering
  \includegraphics[width=\textwidth]{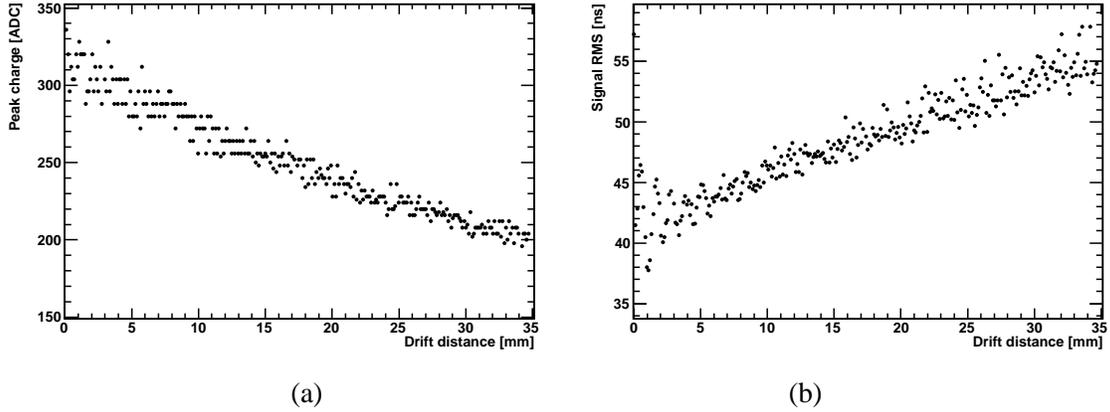} \\ 
  (a) \hspace{6cm} (b)
  \caption{Fine position scanning performed with $V_{\rm gap}=8$~V. (a)~Charge peak
values vs drift distance. (b)~RMS along time-bins extracted from Gaussian fit to the charge
signal vs drift distance.}
  \label{fig:peakrms}
\end{figure}

The positions corresponding to the center of the \emph{plateau} between two consecutive
metallizations have been selected for the following analysis, in order to
minimize possible biases due to laser reflection effects. 
In figure~\ref{fig:peakrms}.a, the charge peak values are plotted as a function
of the drift distance, corresponding to the known laser positions during the
scanning. A decrease of the charge peak value with increasing drift distance is
observed. It is due to the diffusion of the electron cloud during the drift,
which causes a decrease of the peak together with an increase of the signal RMS.
This is confirmed by figure~\ref{fig:peakrms}.b, where the RMS values extracted
from a Gaussian fit to the charge signals along time bins are plotted as a function of the drift
distance.

\begin{figure}[!ht] 
  \centering
  \includegraphics[width=0.32\textwidth]{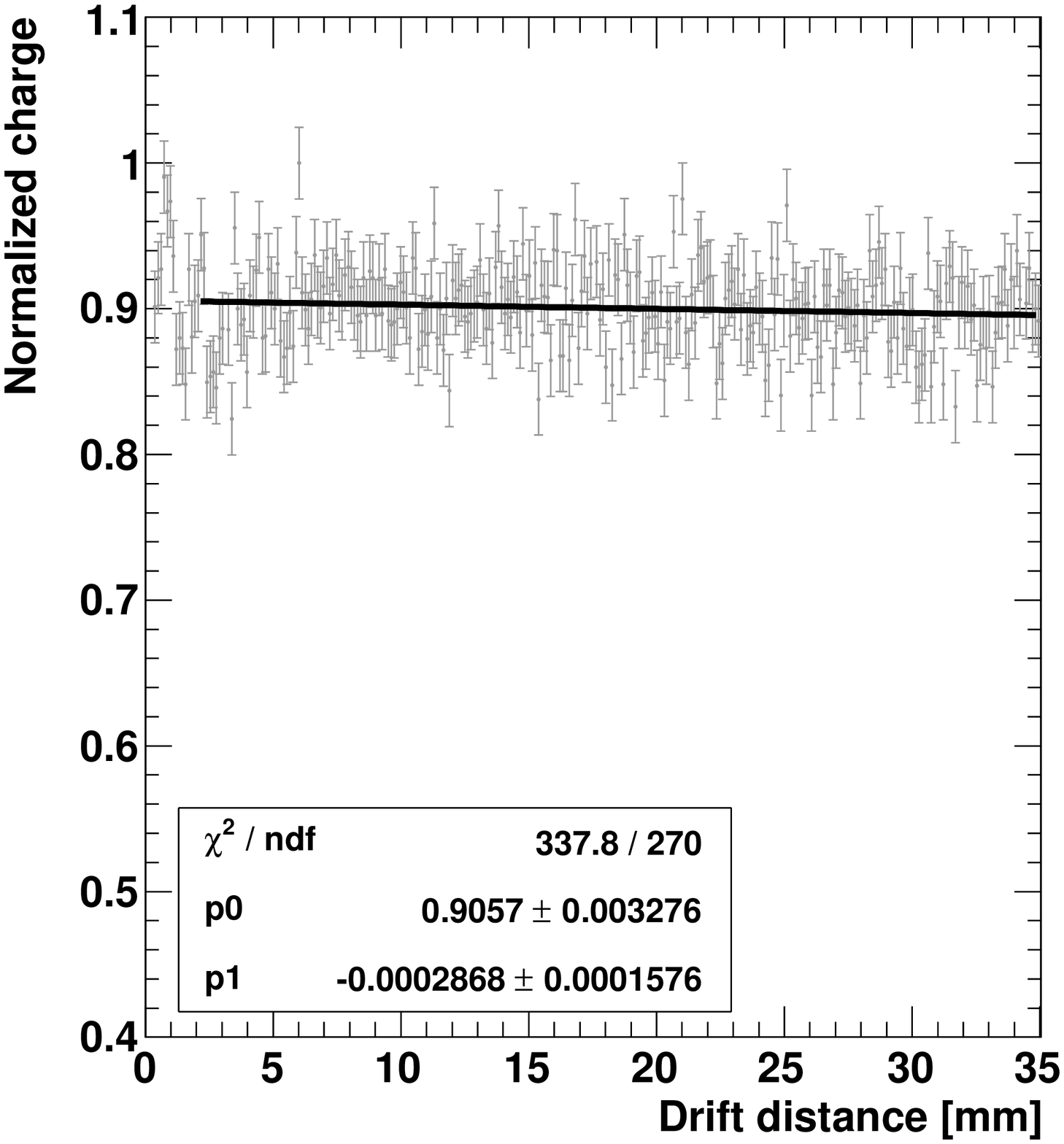} 
  \includegraphics[width=0.32\textwidth]{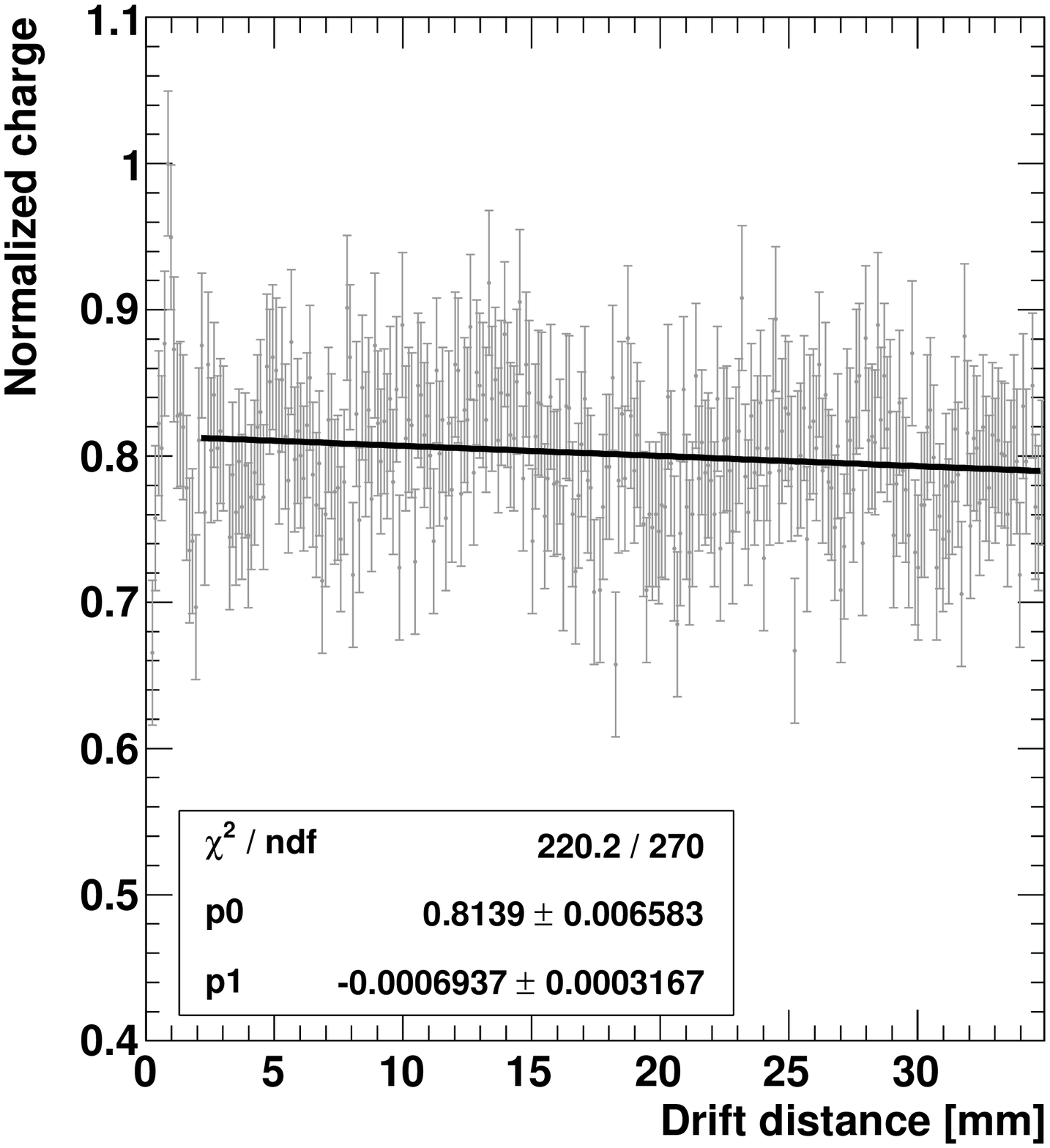}
  \includegraphics[width=0.32\textwidth]{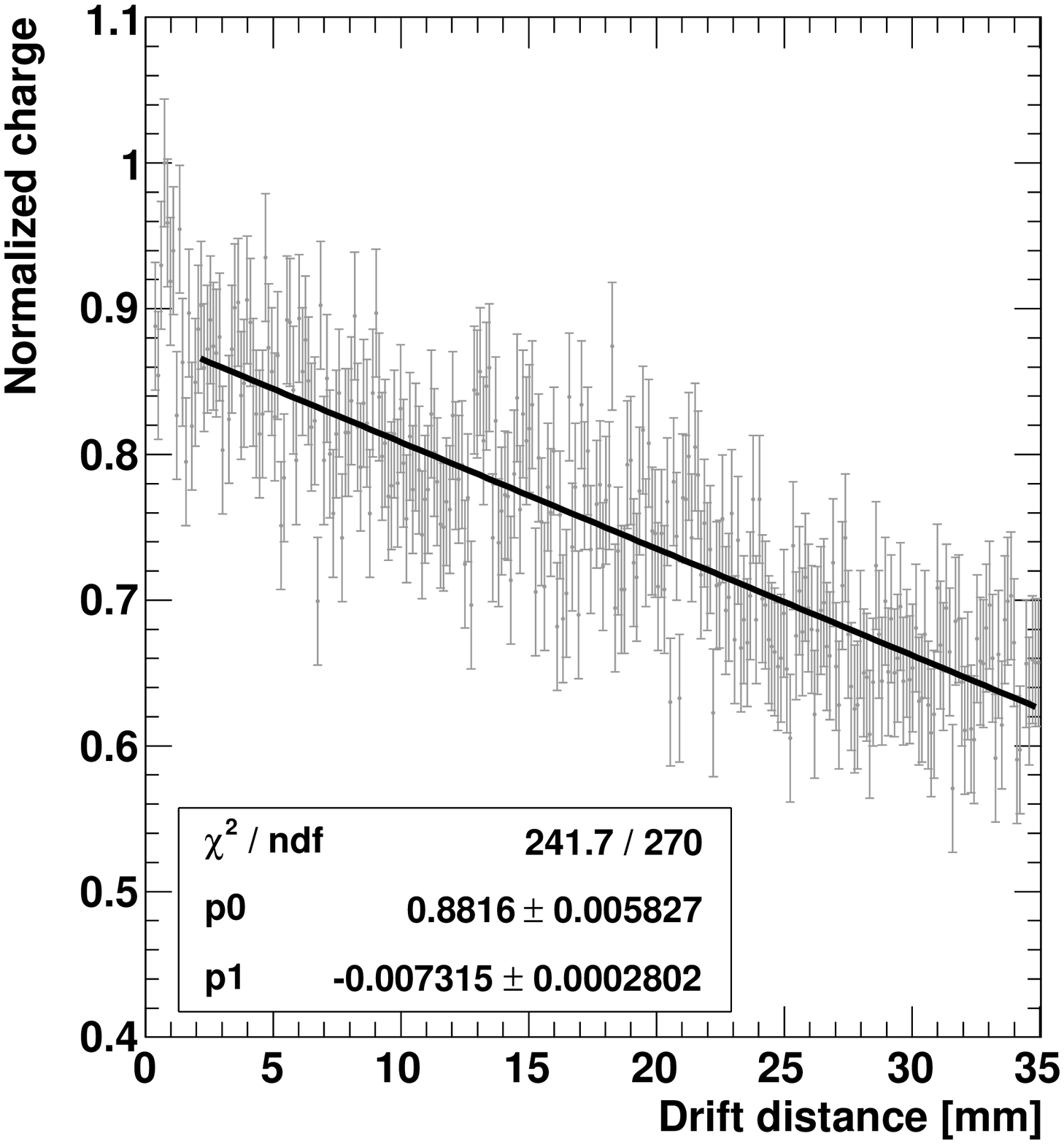} \\
  (a) \hspace{4cm} (b) \hspace{4cm} (c)
  \caption{Collected charge vs laser drift position for three modules
    A, B and C.}
  \label{fig:CL11}
\end{figure}

In figure~\ref{fig:CL11} the total collected charge calculated with the 
two-rectangle method described in \S~\ref{par:method} and normalized to the
maximum value is plotted as a function of the drift distance (corresponding to
the known laser position) for the three modules. The error bars have been calculated
from the RMS of the distribution of the counts summed over the \emph{baseline
rectangle}. Data have been fitted to a straight line of equation $c(x)=a+bx$.
The first 2~mm ($\approx10$~time bins) of drift distance have not been taken
into account for the fit, because it has been observed that at low time bins a strong
effect of common mode noise (i.e. coherent fluctuations of all electronic
channels) appears. It is probably induced by the laser generation
when the trigger signal is issued by the motor control.

For the module A, displayed in figure~\ref{fig:CL11}.a, the collected charge is
independent of the drift distance. The \emph{p1} parameter of the linear fit is
compatible with zero within~$2\sigma$ and the maximum charge difference,
calculated as the line slope multiplied by the maximum drift length (35,085~mm),
is about 1\%. In figure~\ref{fig:CL11}.b (module B) the point-to-point charge
fluctuations are larger than in the previous plot, because the module is
affected by higher noise. The maximum charge difference is about
2\%. The data for these two modules can also be fitted to a constant with a good
$\chi^2$~test value. In figure~\ref{fig:CL11}.c (module C) a charge dependence on the
drift distance is observable, with a maximum charge difference of about 26\%. 
Module C, as mentioned in~\S~\ref{par:method}, is affected by large
inhomogeneities in the Silicon dopant distribution.

The same procedure was repeated for different voltage configurations, in
order to search for possible dependence on the electrical drift field values.
The chosen configurations are summarized in table~\ref{tab:HV}. The nominal
configuration used by the SDD modules mounted in the ALICE ITS is $V_{\rm
gap}=6.04~V$. This value has been chosen as a compromise between the
highest signal peak over noise which is obtained at $V_{\rm gap}=8~V$ and the
better spatial resolution at lower $V_{\rm gap}$ provided by the larger cluster size
due to diffusion effects.
\begin{table}[!ht]
  \centering \small
  \begin{tabular}{|c|c|c|c|}
    \hline
    $HV$ [V] & $MV$ [V] & $V_{\rm gap}$ [V] & $E_{\rm drift}$ [V/cm] \\
    \hline
    \hline
    -2368    & -40      & 8      & 667  \\
    -2082    & -45      & 7      & 583  \\
    -1791    & -45      & 6      & 500  \\
    -1645    & -45      & 5.5    & 458  \\
    \hline
  \end{tabular}
  \caption{Voltage configurations used for the systematic study.}
  \label{tab:HV}
\end{table}

\begin{figure}[!ht]
  \centering
  \includegraphics[width=0.49\textwidth]{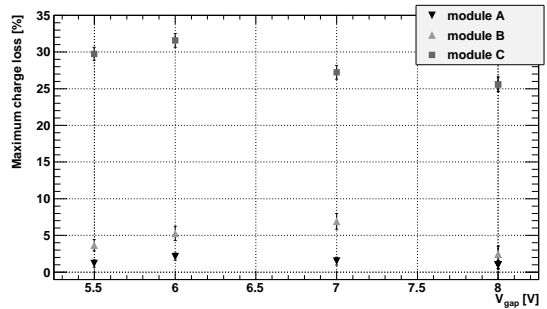}
  \caption{Maximum charge difference values vs voltage configuration.}
  \label{fig:CLHV11}
\end{figure}

In figure~\ref{fig:CLHV11}.a the maximum charge differences for all the three
modules as a function of $V_{\rm gap}$ are plotted. The charge difference values
for module A are compatible with zero, allowing to conclude that no dependence
of collected charge on drift time is present for all the applied $E_{\rm
drift}$. For module B a maximum charge difference of $\approx$5\% for all
voltage configuration is observed. In this case, the CCE seems to be independent
of $V_{\rm gap}$. Module C data present a decrease from 29\% to
26\% with increasing $V_{\rm gap}$. For this particular module, a
significant charge loss during the drift is
observed and this loss is larger at lower values of drift speed, which correspond to
larger drift times.

\subsection{Fixed position}
A second method for measuring the CCE has been developed in order to limit the 
systematic effects due to possible misalignments between the sensor plane and
the laser support structure that can affect the scanning method described in the
previous section. It consists in collecting different samples of
1500 laser shots in a fixed position with different values of $V_{\rm gap}$,
i.e. different values of the drift field. Since the drift time depends on the drift
field as $t_{\rm d}=x_{\rm d}/(\mu_{e} \, E_{\rm d})$, the measurement at
different $V_{\rm gap}$ allows to study the dependence of the collected charge
on the drift time without moving the laser spot on the detector surface. In
order to limit systematic effects due to a drift in the laser intensity with
time, or to possible displacements of the sensor caused by mechanical
vibrations, the measurements were performed with a trigger rate of 100 Hz, thus
limiting to 15 seconds the time for each sample at a fixed $V_{\rm gap}$. 
For each position up to six values of $V_{\rm gap}$ in the range between 5.5 and
8~V were scanned, corresponding to a total measurement time for a given position of
about 3 minutes, including the time to set up the system in between two
measurements and the data acquisition system starting and stopping times. Moreover, the $V_{\rm
gap}$ values were scanned in randomized order, thus canceling correlations
between the drift field value and the time of the measurement.

For each event, the peak value (i.e. the ADC~counts over the baseline in the
anode/time~bin cell with maximum charge), the position and the RMS of the laser
signal along time bins and the total charge were extracted. 
The total charge has been obtained by subtracting the counts
in the \emph{baseline rectangle} from the counts in the \emph{signal rectangle}, 
as explained in~\S~\ref{par:method}. 

\begin{figure}[!ht]
  \centering
  \includegraphics[width=0.49\textwidth]{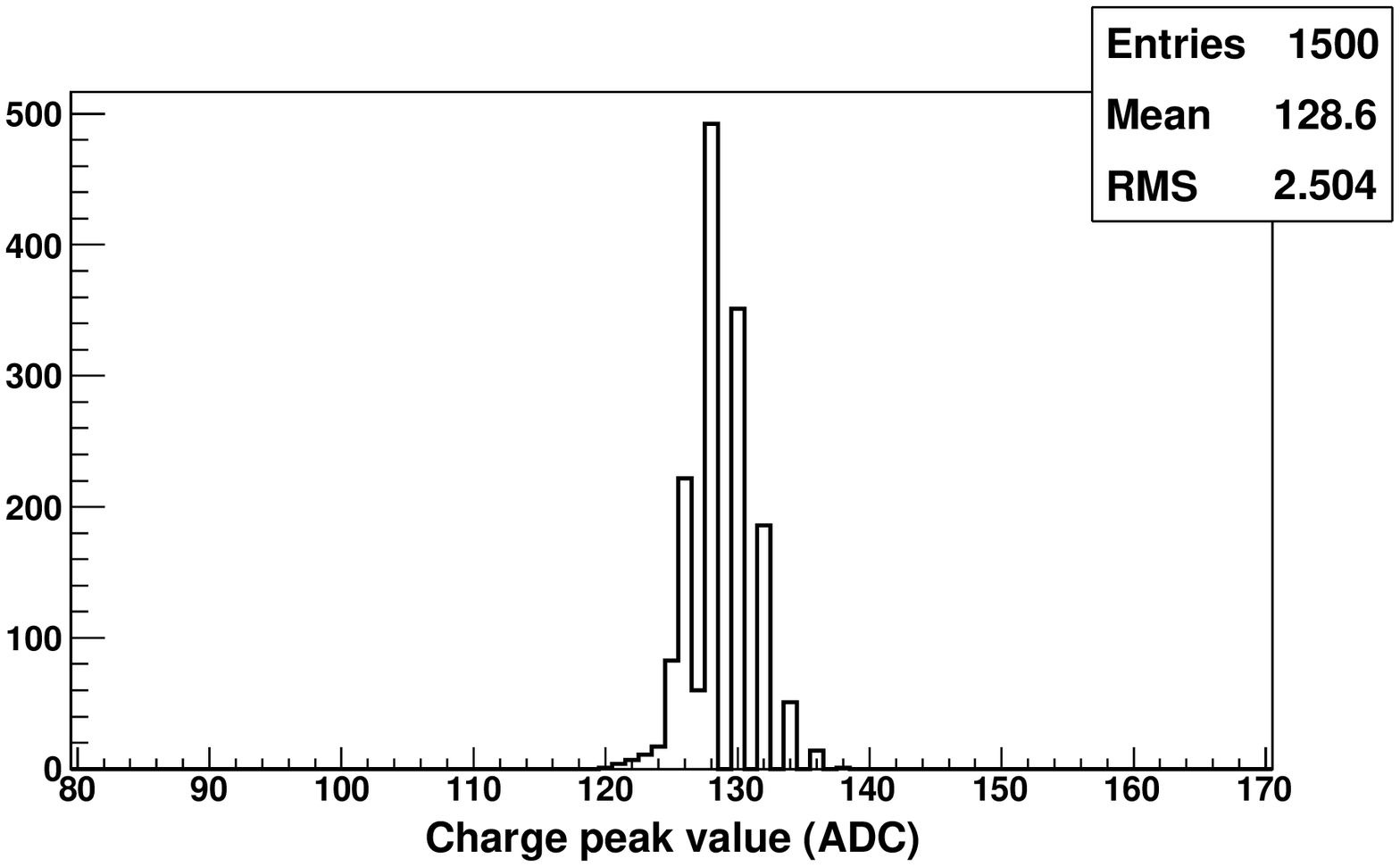}
  \includegraphics[width=0.49\textwidth]{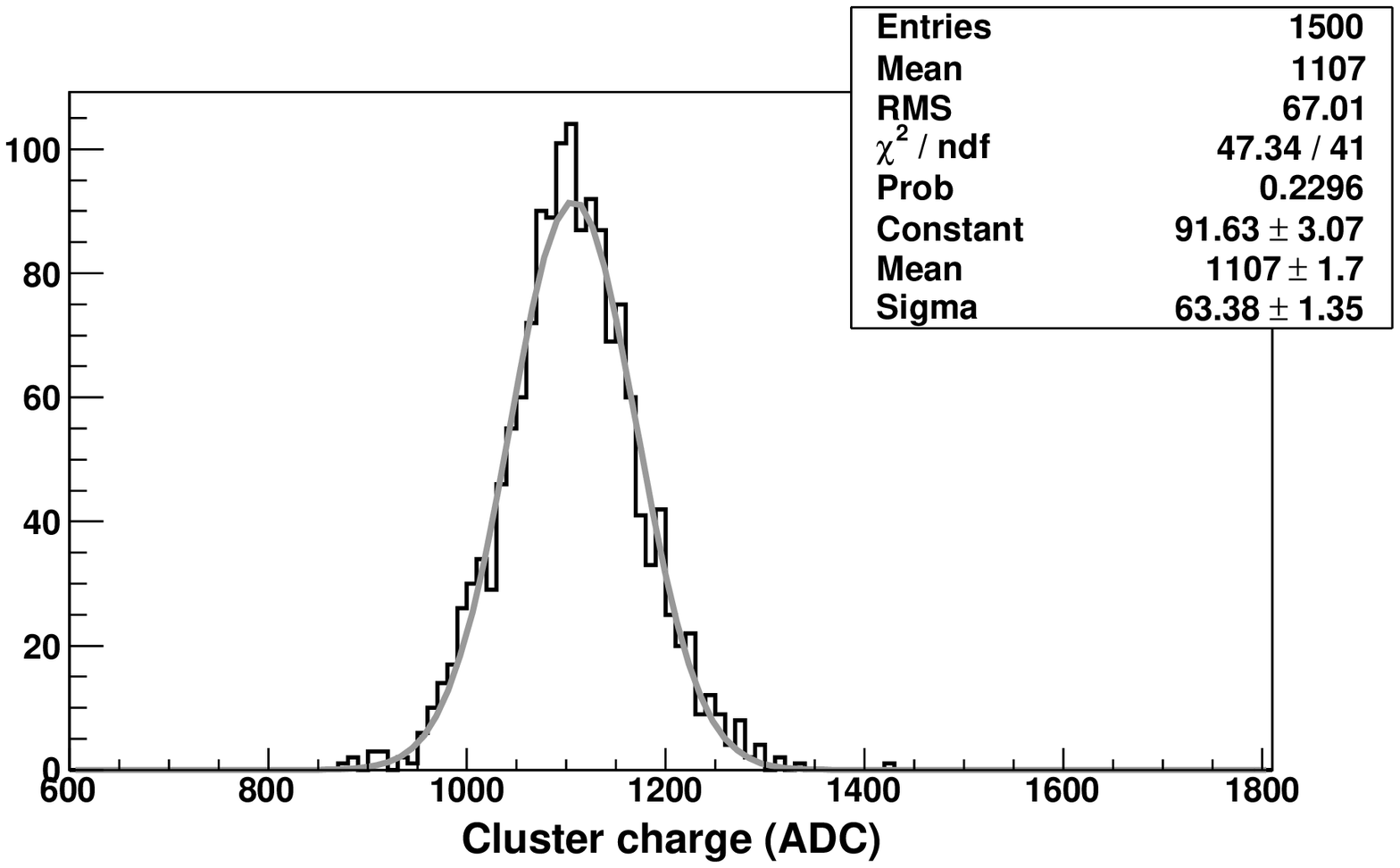} \\
  (a) \hspace{6cm} (b)
  \caption{Distribution of peak charge value (a) and total cluster charge (b)
for 1500~events with $V_{\rm gap}=7$~V.}
  \label{fig:distribution}
\end{figure}

In figure~\ref{fig:distribution}.a the distribution of the charge peak values
for 1500 events collected at $V_{\rm gap}=7$~V is shown. The effect of the
lossy compression from 10 to 8 bit applied in AMBRA is visible: above 128
counts the Less Significant Bit (LSB) is dropped and the precision of the
ADC~counts is limited to 2~units. In figure~\ref{fig:distribution}.b the
distribution of the total collected charge is shown together with the result of
a fit to a Gaussian function. 

\begin{figure}[!ht]
\centering
\resizebox{0.32\textwidth}{!}{%
\includegraphics*[]{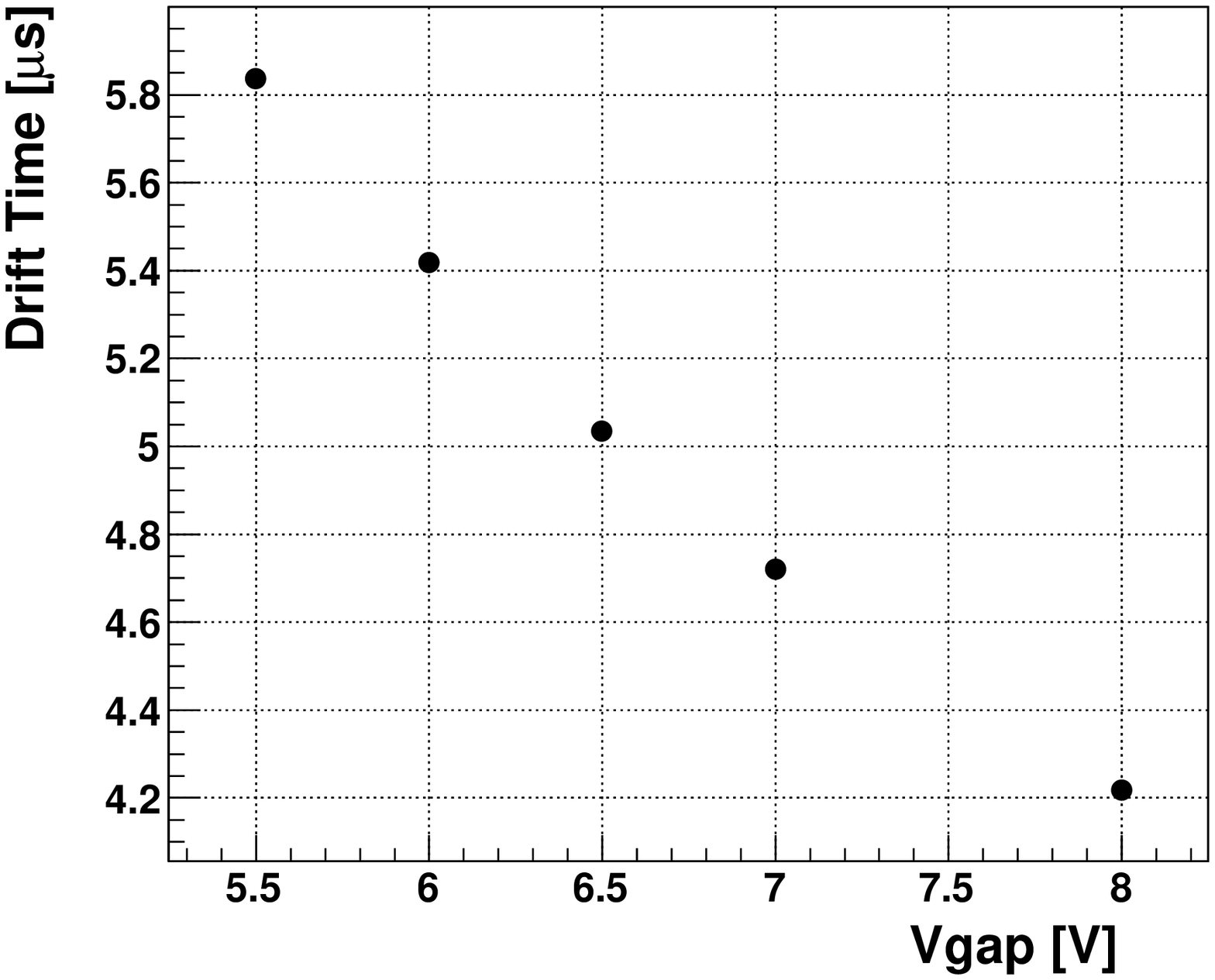}}
\resizebox{0.32\textwidth}{!}{%
\includegraphics*[]{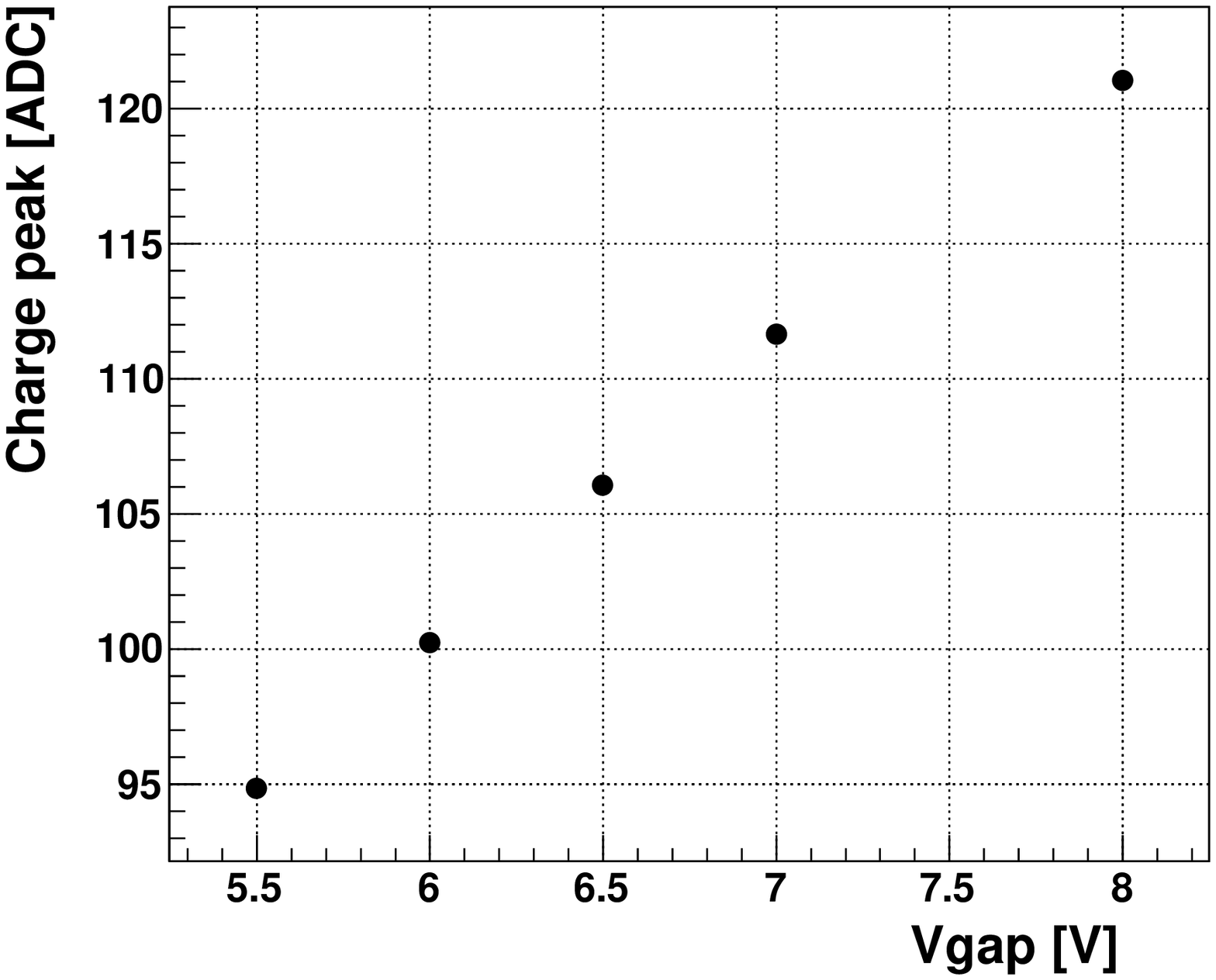}}
\resizebox{0.32\textwidth}{!}{%
\includegraphics*[]{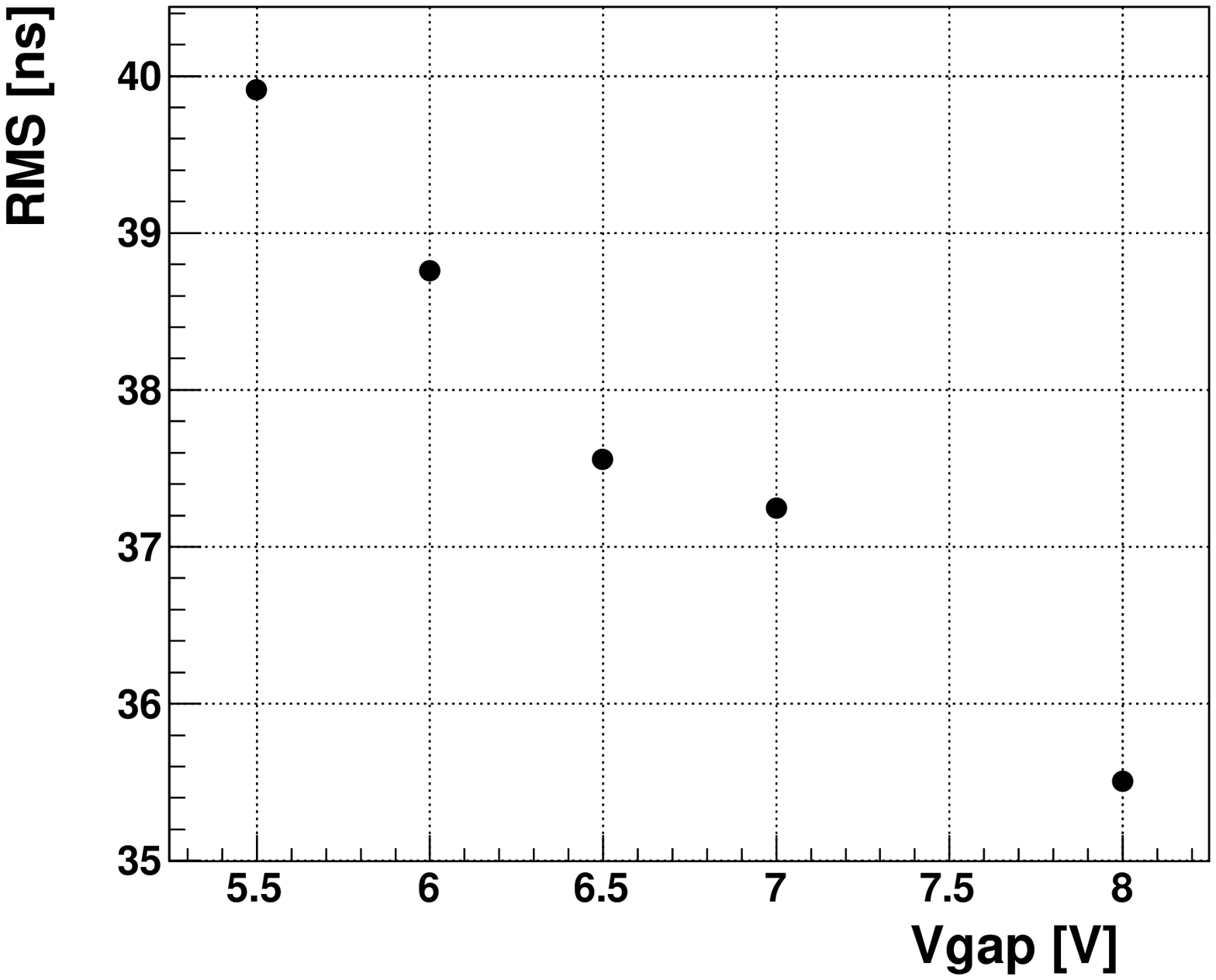}}
(a) \hspace{4cm} (b) \hspace{4cm} (c) 
\caption{Drift time (a), charge peak (b) and RMS along drift direction (c) vs
$V_{\rm gap}$.}
\label{fig:FxPHV}
\end{figure}

The drift time averaged over the 1500 events is reported in
figure~\ref{fig:FxPHV}.a as a function of the applied $V_{\rm gap}$ for a given
fixed position of the laser shot. As expected, a linear decrease of drift time with
increasing drift field is observed. In fig~\ref{fig:FxPHV}.b the average value
of peak charge as a function of the applied $V_{\rm gap}$ is shown: the value of
peak charge increases when the drift field increases, as expected due to the
smaller diffusion of the electron cloud during the shorter drift time. This is
confirmed by the decrease of the RMS of the signal peak along the drift
direction with increasing $V_{\rm gap}$, as it can be seen in
fig~\ref{fig:FxPHV}.c.

\begin{figure}[!ht]
\centering
\resizebox{0.49\textwidth}{!}{%
\includegraphics*[]{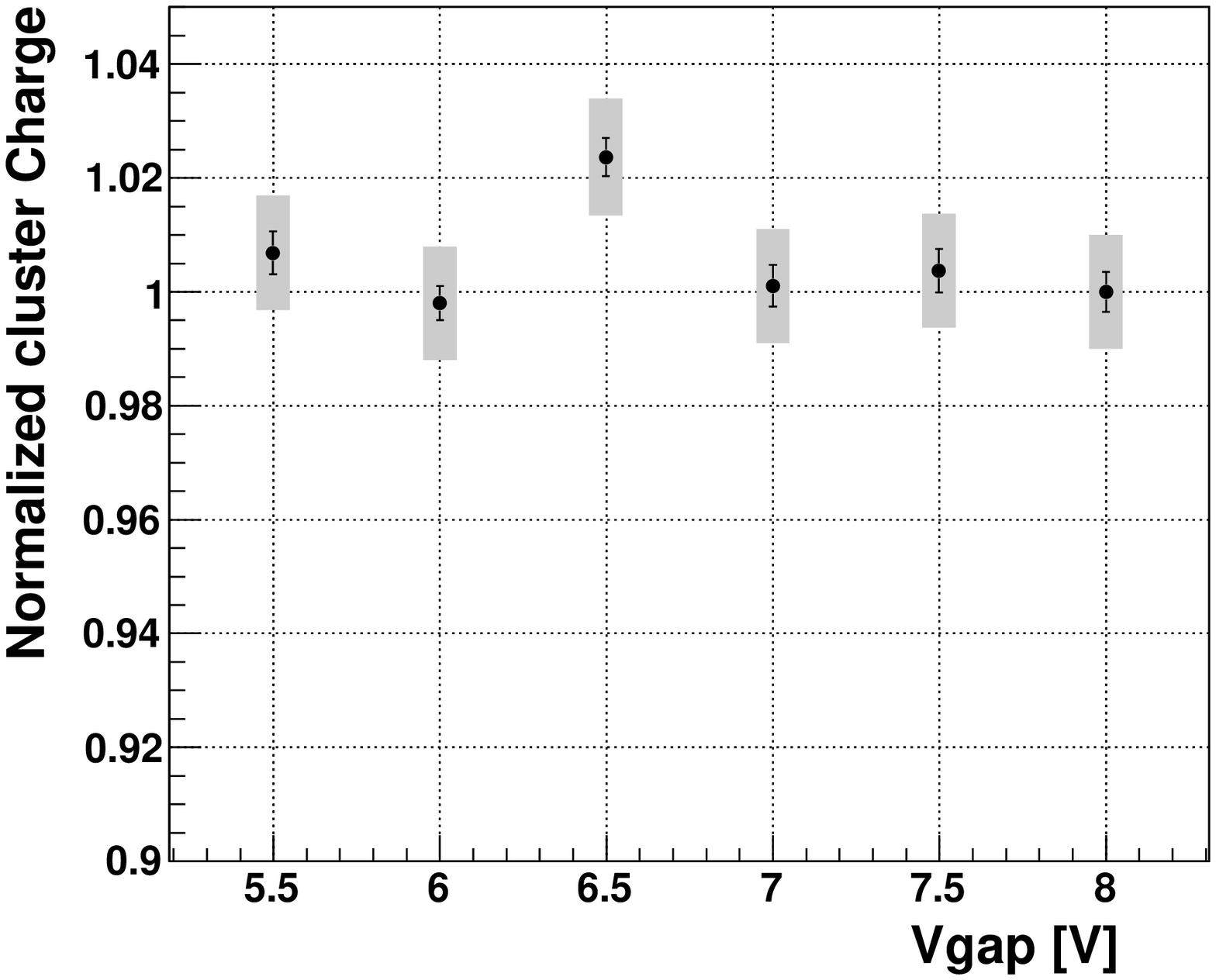}}
\resizebox{0.49\textwidth}{!}{%
\includegraphics*[]{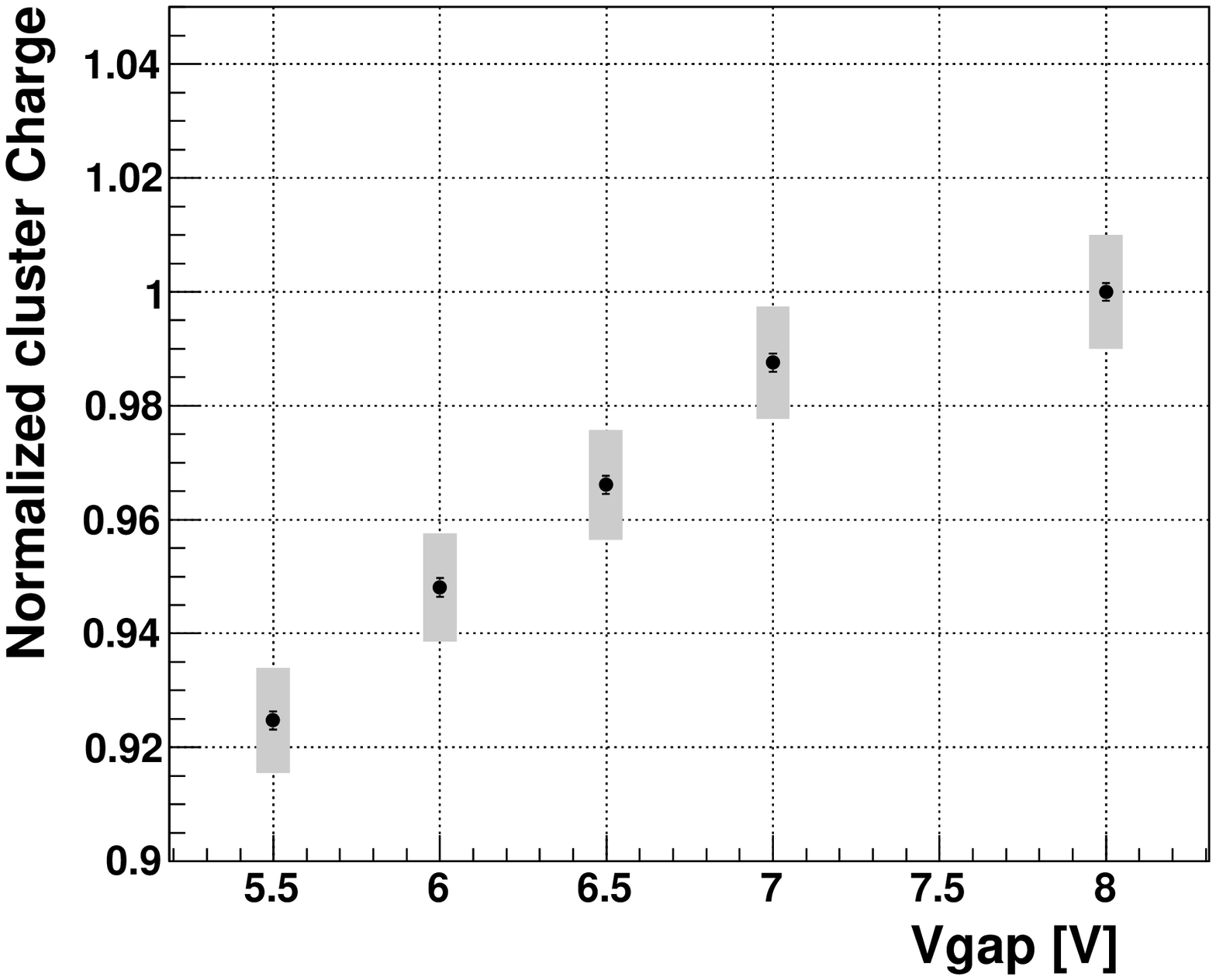}}
\caption{Total collected charge versus $V_{\rm gap}$ for module A (left) and
module C (right).}
\label{fig:integral}
\end{figure}

The values of the total charge averaged over the 1500 events are plotted as a function
of $V_{\rm gap}$ in figure~\ref{fig:integral} for the modules A (left panel) and
C (right panel) used in this study. In order to allow comparison of charge
collected on different modules with different laser intensities, the total
charge has been normalized to the value (in ADC counts) measured at $V_{\rm
gap}=8$~V. The systematic error (gray band around the point) was estimated from
the spread of the values of average cluster charge obtained repeating 6 times
the measurement at a fixed position and at same $V_{\rm gap}$. A clear dependence
of the collected charge on the applied $V_{\rm gap}$ (i.e. on drift time) is observed
for module C, while for module A the total charge is independent of the
drift field, thus confirming the results obtained with the scanning technique
described in the previous section. 

The measurement has been performed shooting the laser in five different
positions on the detector surface. For module A, all the measurements showed no
dependence of the collected charge on the applied $V_{\rm gap}$. For module C, in all
the tested positions, it has been observed that the charge collected at the
minimum drift field ($V_{\rm gap}=5.5$~V) is about 90\% of the value measured at
the maximum $V_{\rm gap}$ of 8~V.

\section{Cosmic rays}\label{sec:cosmic}
A third method to study the CCE has been implemented. It is based on
the measurement of the charge released by atmospheric muons in the SDD
sensor. Using this method, which exploits the signal from charged
(ionizing) particles, it is possible to avoid the possible systematic
effects due to the laser reflection on the metallizations and to
the common mode noise coming from the laser generation triggered by
the motor controller.

\subsection{Trigger system for atmospheric muons}
To study the collected charge at different drift positions in a SDD module with
minimum ionizing particles, a dedicated trigger system detecting atmospheric
muons has been built and operated. The system is composed of three plastic
scintillators, a NIM crate for electronic devices and an acquisition system.
Two of the scintillators are NE102A type with an area of 80~x~80~cm$^{2}$ and a
thickness of 4~cm. The two scintillators are located one meter above and one
meter below the SDD module respectively. Both of them cover the entire area of
the module, but for technical reasons the centers were not aligned to the
center of the module.
The third is a 1~cm tick scintillator with an area of 7.5~x~7.5~cm$^{2}$.
It is located above the SDD module, at 3~cm from the sensor surface.
Due to mechanical constraints, it is not perfectly aligned to the SDD module and
it does not cover its entire surface.

Each scintillator is shielded and equipped with a photo-tube that sends an
analog signal to the electronic system. The analog signal is first discriminated
and then sent to a coincidence unit for trigger purpose. 
The calibration of each scintillator is performed triggering with the two large
scintillators, sending the analog signal of the scintillator to an ADC and
looking at the single particle spectrum in ADC channel units. The
discriminator threshold has been set to 20~mV for all the three
signals and the voltage of the three photo-tubes has been set to
$\approx$1700~V. A coincidence among these three scintillators in a time window
of 100~ns gives the trigger signal to the SDD acquisition system.      

With this trigger, atmospheric muons with direction between 0$^{o}$ and 30$^{o}$
with respect to the vertical direction are selected, providing an average trigger rate
of about 0.2~Hz. Due to the geometrical arrangement of the system and some
selection cuts on the cluster reconstruction in the SDD, 50\% of the triggers	
select a muon crossing the SDD module that can be analyzed. 

\subsection{Results}
\begin{figure}[!ht]
  \centering
  \includegraphics[width=\textwidth]{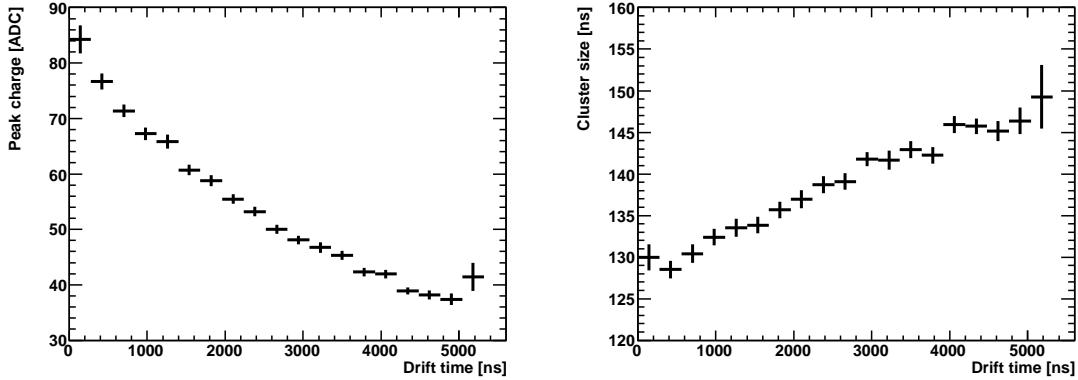}
  \caption{Cosmic-ray data collected with $V_{\rm gap}=8$~V. (a)~Charge peak
    values vs drift time. (b)~Cluster size values vs drift time.}
  \label{fig:cosmicpeak}
\end{figure}

Using the trigger system described in the previous paragraph, we collected about
 40k cosmic events on SDD modules A and C, polarized with $V_{\rm
gap}=8$~V. 
In order to study the effect of the zero-suppression, for module A,
which did not show charge loss effects in the laser-based studies, a
sample of muons was collected with the zero suppression active.
In figure~\ref{fig:cosmicpeak} the results for module A are
 shown (zero suppressed data). In the left panel, a profile of the
 peak charge values shows the decrease of peack charge with increasing of
 the drift time. The maximum decrease is about  47\%. In the right
 panel it is possible to observe that the cluster size increases at larger
 drift times (profile plot). This behavior is consistent with the
 results obtained using the laser 
 methods (see figures~\ref{fig:peakrms} and~\ref{fig:FxPHV}). 

To study the dependence of the collected charge on the drift time, data have
been acquired without zero suppression and, for each event, the charge of the
cluster has been calculated with the method described in \S~\ref{par:method}. 

\begin{figure}[!ht]
  \centering
  \includegraphics[width=\textwidth]{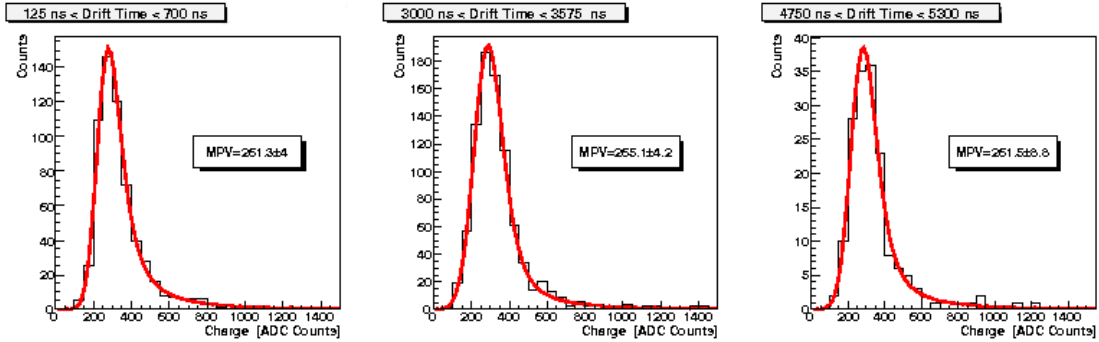}
  \caption{Distributions of collected charge in different drift time intervals,
fitted to a convolution of a Landau and a Gaussian (without zero suppression).}
  \label{fig:landau}
\end{figure}

Data have been divided in nine bins along the drift direction. For each bin the
distribution of the collected charge has been fitted with a convolution of a
Landau and a Gaussian. As an example, three of these distributions, one for a
bin close to the anodes, one for the central region and one for a time interval
far from the anodes (i.e. close to the center of the detector) are shown in
figure~\ref{fig:landau}.

\begin{figure}[!ht]
  \centering
  \includegraphics[width=0.49\textwidth]{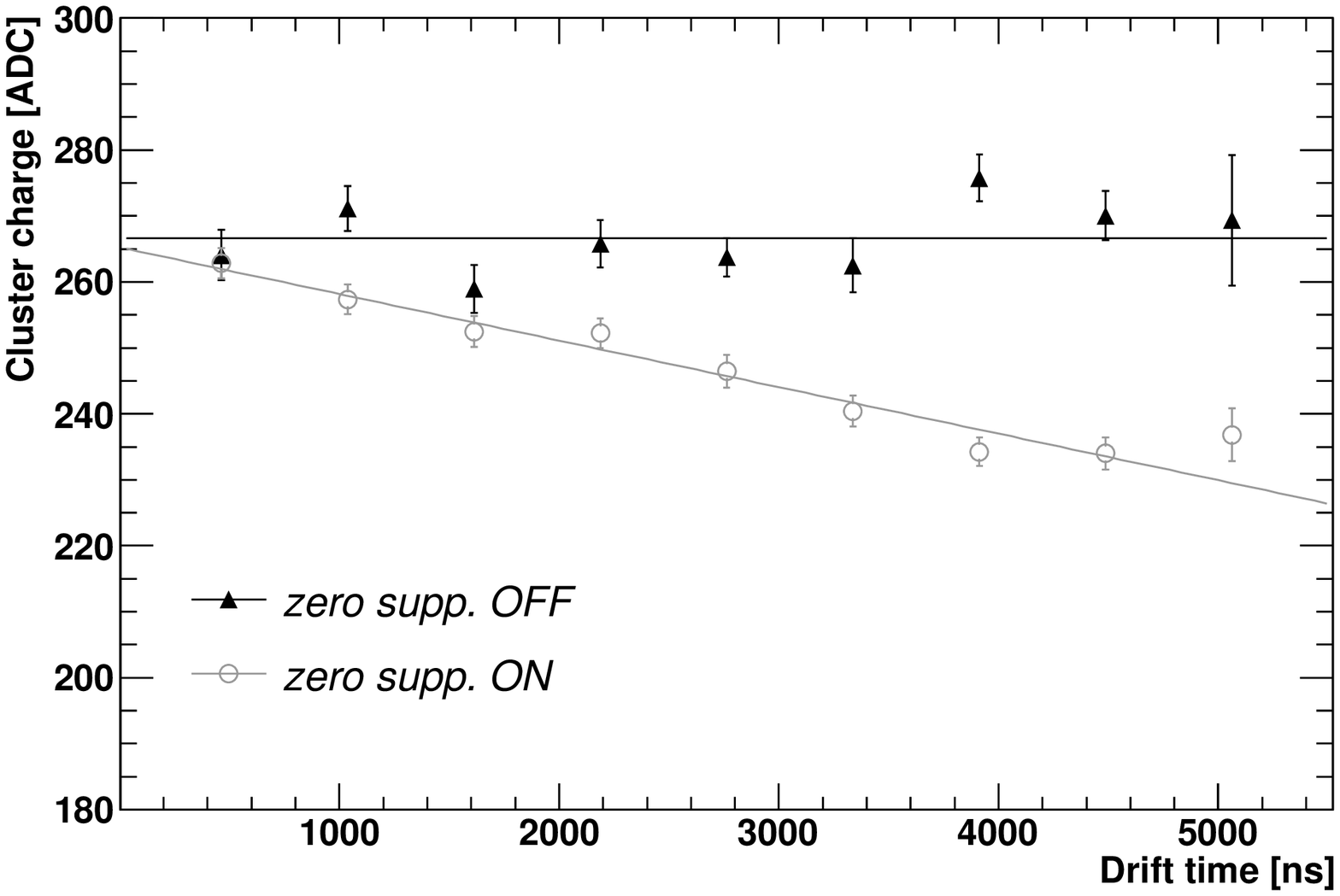}
  \includegraphics[width=0.49\textwidth]{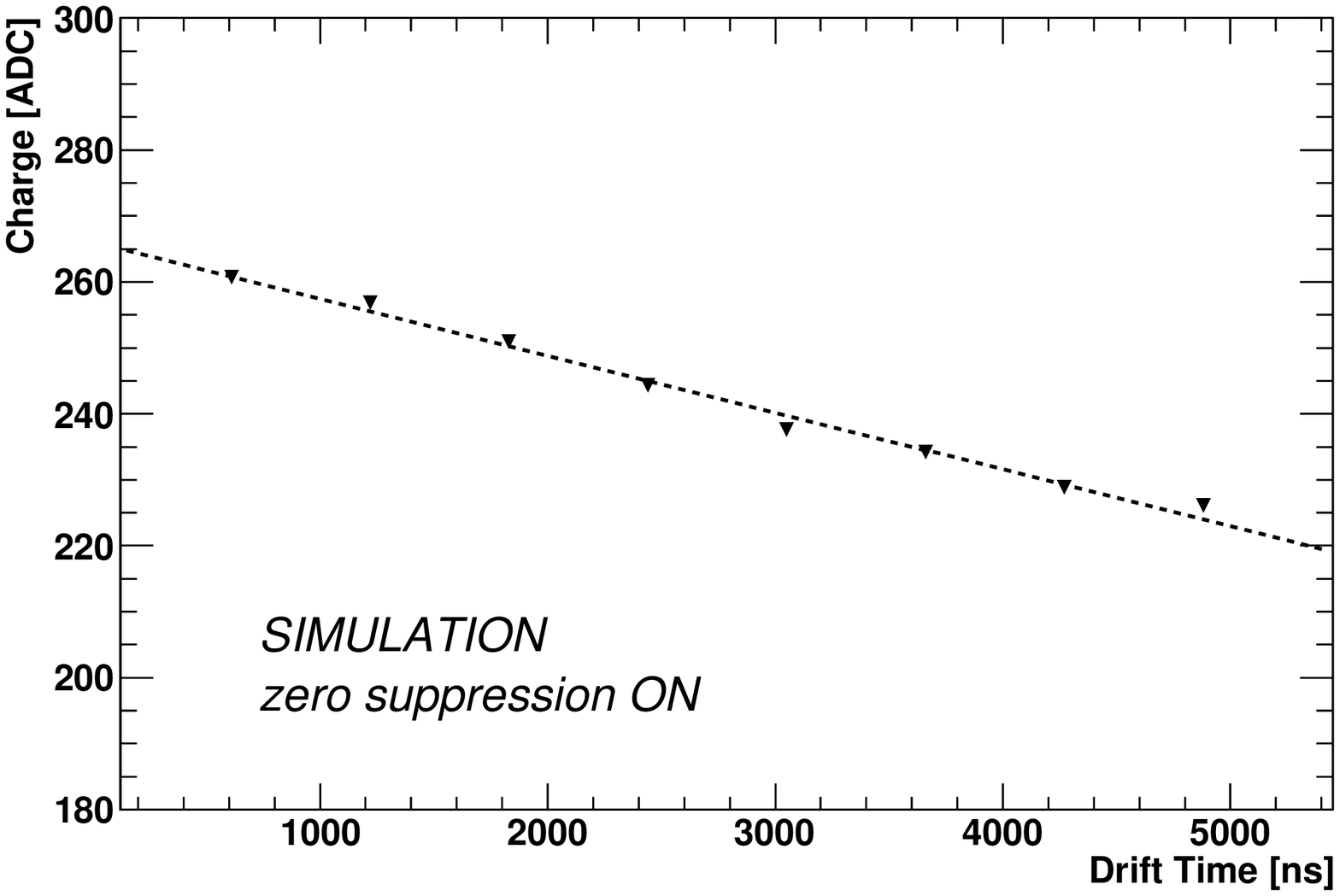}\\
  (a) \hspace{6cm} (b)
  \caption{(a) module A: MPVs versus drift time. Data taken without zero-suppression
    are fitted to a constant, data taken with zero-suppression on are
    fitted to a straight line. (b) Monte Carlo simulation: cosmic data
    taken with zero-suppression.}  
  \label{fig:cosmic11}
\end{figure}

The Most Probable Value (MPV) of the fit functions is then used in the study of
charge dependence on drift time. In figure~\ref{fig:cosmic11}.a the MPVs as a
function of the drift time, for module A, are plotted
for two sets of data (with and without zero-suppression). The
triangular markers represent data acquired without zero-suppression in the
CARLOS chip and they do not show a dependence of the collected charge on drift
time. They can be fitted to a constant function.
The circular markers represent data collected with zero-suppression and have
been fitted to a straight line. In figure~\ref{fig:cosmic11}.b, the
MPVs as a function fo the drift time obtained by a Monte Carlo
simulation, including a detailed description of the detector and
front-end response, are shown and fitted to a straight line. In case
of zero-suppressed data, the difference between charge collected for
muons crossing close to the
anodes and muons with maximum drift distance amounts to $\approx$15\% in case
of data and to $\approx$17\% in case of simulation. This confirms
that the simulation correctly reproduces the detector response and the
combined effect of charge diffusion and zero suppression on the
collected charge, allowing to use a correction factor extracted by the Monte Carlo simulation in the
offline data reconstruction. 

\begin{figure}[!ht]
  \centering
  \includegraphics[width=\textwidth]{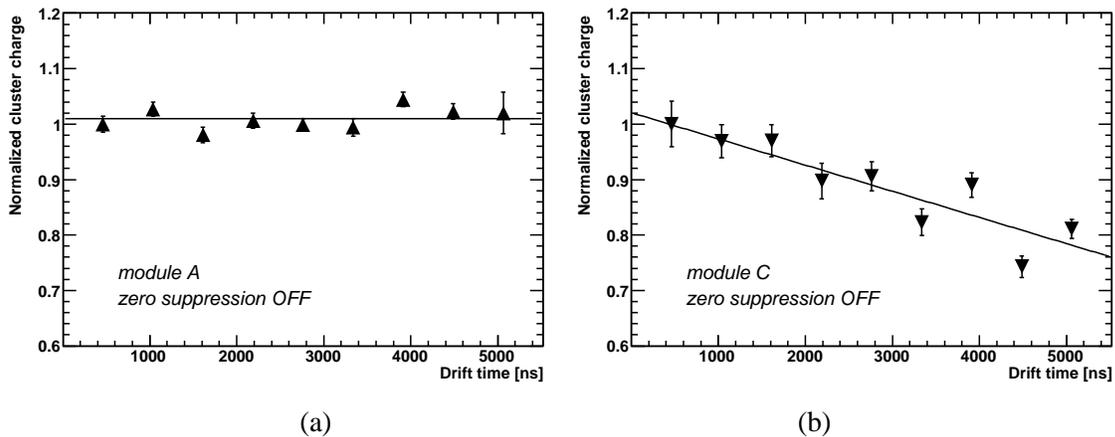}\\
  (a) \hspace{6cm} (b)
  \caption{MPVs versus drift time (normalized values). (a) Module A,
    data fitted to a constant. (b) Module C, data fitted to a straight line.}  
  \label{fig:cosmic2}
\end{figure}

In figure~\ref{fig:cosmic2}, the MPVs extracted from
non-zero-suppressed events are plotted as a function of the drift
time, for modules A and C.
Data are normalized to the value at the lowest drift time.
For module A (left panel), data are fitted to a constant. As seen
in the previous plot, no decrease of the most probable value of the
collected charge is present as a function of the drift time.
On the contrary, for module C (right panel), a cluster charge dependence on
the drift time is visible. Data have been fitted to a straight
line. The maximum charge loss
value, extracted from the fit, is $\approx$26\%. 
This result is compatible with to the one obtained with the scanning
methods using the infrared laser.

\section{Conclusions}\label{sec:conclusion}

A systematic investigation of charge collection in the Silicon 
Drift Detectors has been performed on three sensors 
spared during the ALICE Inner Tracking System construction.
The CCE has been investigated by studying the total cluster charge
as a function of drift time/distance for signal events produced
with an infrared laser and for atmospheric muon clusters.
On modules with uniform dopant concentration (A and B), no dependence
of the collected charge on the drift time has been observed, allowing to conclude
that effects of electron trapping during the drift are negligible.
On the contrary, on modules with large dopant inhomogeneities a
significant decrease of collected
charge with increasing drift time is observed, reaching a  $\approx$~26\%
difference
between clusters produced close to the anodes and clusters produced
in the center of the detector.
It should be pointed out that for this kind of sensors (C),
an inefficiency in charge collection was already
measured during beam tests~\cite{beam} and the significant inhomogeneities 
in dopant concentration were observed when mapping the detector response
with the laser~\cite{articolomappature}.
As a matter of fact, only 2 modules of the 260 that have been mounted on 
the ALICE Inner Tracking System have been built on the particular
wafer type (C) and are expected to 
be affected by sizable systematic effects on charge collection efficiency.
 
It has also been shown that the zero-suppression algorithm applied to
reduce the data size affects the measured cluster charge in a way that depends
on the drift time: for larger drift times, the electron 
diffusion gives rise to wider cluster tails that are more likely to be cut
by the thresholds applied when suppressing the zeroes.
It should be pointed out that this effect can be accounted for when correcting
the reconstructed cluster charge because it is quantitatively reproduced by 
detailed Monte Carlo simulation of the SDD detector response. This
correction is possible because, also at large drift times, the peak charge
values are higher than $\approx$20~times the average noise, as is it
shown in figure~\ref{fig:cosmicpeak}.

\acknowledgments
We wish to thank the people helping us in setting up the 
mechanics and electronics of the SDD test station:
F.~Borotto, F.~Cotorobai, M.~Mignone and F.~Rotondo. 
We wish to thank also the people involved in the assembly and bonding
of the SDD modules, whose work was fundamental for the SDD project: F.
Dumitrache, B. Pini and the teams from SRTIIE, Kharkov, the Ukraine.
This work was partly supported by the European Union, 
the \textit{Regione Autonoma Valle d'Aosta} and
the \textit{Ministero del Lavoro e della Previdenza Sociale}.

\end{document}